# A Conceptual Design of Mobile Learning Applications for Preschool Children

Radoslava Kraleva[#], Velin Kralev[#], Dafina Kostadinova[*]

[#] Department of Informatics, South West University "Neofit Rilski", Blagoevgrad, Bulgaria

[*] Department of Germanic and Romance Studies, South West University "Neofit Rilski", Blagoevgrad, Bulgaria

*Abstract*— **This article focuses on the possibilities of using mobile learning in the Bulgarian preschool education of young children. The state preschool educational regulations are presented and discussed. The problem concerning the children's safety when using mobile devices in terms of access to information on the Internet is revealed and analyzed. Two conceptual models of applications for mobile learning aimed at preschool children are designed. Their advantages and disadvantages are outlined and discussed.**

## I. INTRODUCTION

Mobile technologies have made significant progress in its development within the recent years. They have become an integral part of people's daily life [1]. Therefore, the great interest in the use of mobile technology in education is easy to understand. It was only a few years ago when the only teaching practices related to computer devices were the distance learning (e-learning) and ubiquitous learning (u-learning). Today it is mobile learning (m-learning) that is in the focus of attention and a number of scientists from all over the world work on its development and distribution. Such research is presented in [2], [3], [4] and in many others.

The term mobile learning denotes the use of mobile applications installed on a mobile device (smartphone or tablet) that are used in the education process at school and at university [5].

The implementation of mobile learning is not a simple process, especially when it is aimed at young children's learning. It is closely related not only with the authors of textbooks, teachers, children (students), but also with the parents who have a crucial influence on the way their children perceive the outer world: they play an important role in the formation of their character, interests and social behavior. The phenomenon of this influence is represented in detail in [6].

The opinion of parents represents an essential part of the process that is associated with the implementation of this new type of learning of young children. In order to obtain an objective assessment of their attitude regarding this subject an inquiry research among parents of children of different ages was made and the results from it were published in [7]. The data obtained from this study revealed that more than 90% of the parents support the idea of using mobile learning. This article came as a natural addition and further elaboration of the analyses and studies made in [7].

To claim that a certain textbook is good and meets the standards it must comply with the State Educational Regulations (SER) for the respective age group. However, these regulations are different for different countries. Therefore, for the purposes of this investigation the SER in Bulgaria must be presented and analyzed.

Last but not least, special attention should be paid to the safety of children when using any computing devices that provide connection to the World Wide Web.

Unlike similar articles that focus on the technical specification of the software for mobile learning [8], here the optimal balance between the quality of the presented model, the skills of the users (mainly children) and the provision of transparent and safe way of learning will be sought.

On this introductory background the main goal of this study may be defined as follows: to present a generalized framework of applications suitable for mobile learning by preschool children, so as the State Education Regulations should be met and the safety of children in terms of access to information on the Internet be provided.

To achieve the goal several tasks have been set in this study:

- Presenting the preschool SER in Bulgaria;
- Discussing the safety of children when using mobile devices;
- Analysis of the interface limitations associated with mobile devices;
- Summarizing the requirements for mobile learning applications for young children;
- Designing a generalized conceptual framework of applications for mobile learning.

## II. ANALYSIS OF STATE EDUCATIONAL REGULATIONS

As mentioned in the introduction, when it comes to developing new learning tools it is important to analyze the State Educational Regulations (SER). Regulations concerning the pre-school education were published for the last time in 2005 in Decree № 4 of 18.09.2000 [9].

Based on this document [9] the following modules can be distinguished (Table 1).

TABLE I.  SER REGARDING THE TEACHING MATERIAL REQUIREMENTS FOR PRESCHOOL CHILDREN IN BULGARIA [9]

| Table Head | Table Column Head |
|---|---|
| *Bulgarian language and literature* | • Introduction into the Bulgarian language.<br>• Continuous speech, understanding the words and the use of the words.<br>• Simple and complex sentences.<br>• Correct pronunciation.<br>• Distinguishing between sentences, words and sounds.<br>• Identify letter graphemes and writing on the dotted lines.<br>• Identification of a sequence of actions based on an image and recognition of literary heroes.<br>• Interpretation of literary works.<br>• Completion of a tale based on an image. |
| *Mathematics* | • Knowledge related to quantitative, spatial and temporal relationships of objects, development of thinking and memory.<br>• Counting, addition and subtraction to 10.<br>• Comparison: more, less and equal.<br>• Units and concepts of latitude, longitude and altitude<br>• Spatial relations related to determining the position of objects relative to each other, solving problems with maze and searching for a route.<br>• Terms related to parts of the day, days of the week and the seasons.<br>• Recognition of geometric shapes, such as square, circle, triangle, rectangle, top, side and others. |
| *Social world* | • Getting acquainted and orientation in the surrounding world.<br>• Communicating with others and self-affirmation.<br>• A subject area.<br>• Healthy and social environment.<br>• Cultural and national values. |
| *Natural world* | • Getting acquainted with the flora and fauna, the nature, the natural phenomena and laws.<br>• Models of behavior in the nature. |
| *Art* | • Basic skills related to creative activities and children's creativity<br>• Focusing on artistic work, application and modeling. |
| *Physical education* | • Physical self-improvement, self-control, drive for motor and personal activity.<br>• Applying one's own ideas and experiences in a variety of conditions. |
| *Music* | • Formation of basic skills and personal approach to music.<br>• Basic knowledge of the art of music. |
| *Constructive technical and household activities* | • Ways to construct a variety of products and fine motor training.<br>• Paper work: folding, cutting and gluing; sewing, tying, interlacement, drawing straight and curved lines, connecting the dots. |
| *Games culture* | • Games environment.<br>• Development of artistic abilities.<br>• Introduction to the Bulgarian folklore. |

The aim of the introduction of these educational requirements in Bulgaria is to outline the needed functionality which an application must have so that it can be used in the education of children at the primary school age.

## III. SAFETY OF CHILDREN WHEN USING MOBILE DEVICES

Another important aspect of the implementation of mobile learning is the child's safety. The notion of 'safety' presupposes limiting their access to the Internet or to other applications that can lead to some negative effects on their health and/or physical and/or mental development. With a connection to the global network external attacks aiming at gaining access to personal data are possible. This issue has been widely discussed in a large number of scientific works, among which [10], [11].

The use of mobile devices including in education and their impact on the social life of young children and teenagers is presented in [12]. What is noted is the fact that their priority is to provide updated information at any time and place. On the other hand, they have a negative impact as well, since their use leads to alienation of the children.

Apart from the influence on the intellectual development of young children, the use of mobile devices can affect children's health as a whole. Such health issues are presented and examined in [13].

Research related to the safe use of various mobile devices in terms of the dangers threatening young children and their parents is presented in [14]. These are massive media advertisements with the purpose of purchasing a commodity; showing any event in the life of the children in the social networks; chatting with strangers and others. Topics such as cyber-bullying, pornography and risk of addiction to mobile devices are taken into consideration. Quotations from interviews with both adults and children are provided, thus, the actual opinion of consumers has been revealed. What is more, the children's preferences of using mobile devices in accordance with the personal and portable computers are investigated and the results have been provided here.

As a result of this research the following conclusion can be drawn: when developing applications for mobile learning for children, the most important is to ensure reliability and limiting external attacks. One solution to this problem is the use of Web services that restrict the access to external Internet resources, except for those that are necessary for the application itself.

## IV. REQUIREMENTS TO THE MOBILE APPLICATIONS FOR TEACHING YOUNG CHILDREN

The topic related to the use of mobile devices as a learning tool is currently important and up to date. Before the model of a single framework for mobile application is presented, first, the criteria which it must comply with must be specified. Similar studies have been already conducted and their results have been supplied in other publications, such as in [15]. However, no clearly defined requirements for the software designed especially for preschool children have been stated there. These studies are mainly aimed at students at the secondary education and university students.

In another article [16] principles for the implementation of mobile learning as a tool for distance learning in the form of web-based applications, using a browser on one's mobile device are presented. However, with this model a direct Internet connection is required.

Based on the statements made in the sections above, several requirements that an application must meet in order to claim that it is appropriate for a mobile learning for young children can be determined:

*1) To provide safe environment for children:* If Internet connection is needed, it can be secured so as to limit children's access to inappropriate content and to ensure the cessation of external attacks to their personal data;

*2) To cover the teaching material stated in SER of the country*;

*3) To be in the native/official language of the children*: in our case, the requirement is to be in the Bulgarian language;

*4) To be with a friendly and intuitive interface* that will allow children to use the application easily.

In [17] a detailed classification of the existing applications for mobile devices that could be used for mobile teaching to preschool children in Bulgaria has been made. To sum it up, the obtained data concerning the existing mobile applications in stores of Google, Apple and Microsoft until the moment of writing this article cannot suggest that the application of modern mobile teaching and learning at preschools can take place. To achieve this goal, it is necessary to develop an appropriate application that is in compliance with the criteria listed in this article.

## V. CHILDREN INTERFACE LIMITATIONS OF MOBILE APPLICATIONS FOR CHILDREN

What is typical of mobile devices is that they have limited resources of the hardware components.

The most significant disadvantage is the size of the screen. Today there are still smartphones with a *screen size* of 2.4" and a resolution of 240x320 pixels. Hence, it results in a serious restriction on the size of the objects that can be displayed as well as in the size of the fonts.

Even the *font* with size 12pt is usually too small and hardly readable even at a larger screen. With such screens the rule is not to use serif fonts, which is contrary to the rules laid down in the prepress related to the texts intended for children. Furthermore, when specific custom fonts are used, it must be certain that they will be readable and render nicely on the most devices.

The used *buttons* must be big enough and must stand out from the rest of the interface part, so that the user can easily navigate in its utilization. The best practice in this case is that the buttons should only contain graphic image. So the text on them will not be incomprehensible and confusing for children.

The child (user) must be always able to return to *the home page*, so that the interface of application might be defined as an intuitive one.

In addition to using *the command buttons*, the application should provide an opportunity to implement voice commands, too. At this stage, however, it would be difficult to do so, because the scientific work related to the recognition of children's speech in Bulgarian are too few [18].

A detailed explanation of the limitations associated with the design of the mobile applications interface can be found in [19] and [20].

An analysis of the interface applications and ways of perception of the advanced mobile technologies by children was made in [21].

## VI. CONCEPTUAL DESIGN OF APPLICATIONS FOR MOBILE LEARNING

Bearing in mind the above rendered facts, it can be inferred that mobile learning is already seen as necessary and great help to overcome the barriers of time and space associated with providers of knowledge. This is how the access to updated information increases. Sometimes the manner of teaching depends on the curricula and on the different ways of building a teaching system. Therefore, when building such systems, the required simplicity of the teaching material must be achieved and the influence of the used technology on the quality of the developed application must be reduced. Furthermore, any system for mobile learning must provide a safe and secure environment for consumers.

There are some publications related to the design of applications for mobile learning. Such are [22] and [23], but they deal with more abstract patterns and differentiate the types of mobile learning depending on the modalities of training and are not aimed at preschoolers.

A variety of design models for mobile learning applications are proposed in the Book of papers [24]. But most of them have become obsolete because of the rapid pace of development of mobile technologies over the past five years.

The education of preschool children can be conducted only by parents, or with collaboration between parents and teachers. This article discusses these two situations as different cases of use of mobile learning. This is necessary in order to cover the various opportunities that arise when developing a mobile application designed for teaching to young children. This will allow the smooth development of real systems. This will allow the unproblematic development of real systems.

The proposed models are based on the best practices that are explored in contemporary literature, on the analyzed parents' opinion in Bulgaria and on the authors' own view on the problem. In these conceptual models the kind of mobile operating system is ignored because modern integrated development environment provides a wide range of cross-platform development tools. Moreover, applications are often designed only to solve a specific task and they have a varying type of interface and architecture. That is why our model seeks to provide flexibility in the development of real prototypes.

It should be noted that in both cases a remote database in which there are teaching materials that can consist of textual and multimedia information is used.

The organization of the application interface is not essential at this stage, as in the modern programming the best practices of design and programming are respected and followed.

*A. Mobile application for individual learning*

The first case is applying mobile learning to children who do not attend school. This is the so-called individual learning and can be done at home by the parent.

An individual mobile application can be used in this case; the parents can download to their phone (tablet) or that of the child. This application must be able to provide a secure access to all its options without a connection to the Internet or some other type of network. Additionally, the application must support two roles modes – as a parent and as a child.

The functions of the role of the parents are available after verification by the user password that was created during the initial installation of the application. Parents should have the right to monitor the full diary of the child when working with the application. With their parental role they can observe the duration of the application usage, the daily and monthly performance achievements, the access control to various modules and functions, and to check for updates of the application. When a new version of the application is available, a second verification with the access password is needed and it is only then that the upgrading will be carried out.

The access to the various functions of the mobile application is represented by the use case diagram, shown in Fig. 1.

The interface of the mobile application designed for individual learning by young children must be designed as a funny game.

The problem that exists in most similar applications is that the parent can only allow or limit certain functions and the update of the educational content is done only by the developer.

What is new in the model proposed in this paper is that the parent can update the content of the application. From a technical perspective it can be done in two ways: through the parent's account or via another application that is installed on a computer device such as a laptop or a desktop computer that a parent can update with the needed content, add new tasks and remove these that are no longer necessary. The activity of the developer is limited to adding new functionality to the application; he/she does not have to update the educational content.

*B. Mobile application for collective learning*

The second case of a mobile application design is based on the idea of collective learning (fig. 2) that involves students, teachers and parents.

The concept here is that everyone has a different application, appropriate for its role, installed on a personal mobile device.

All applications have access to a remote database server with a common database. It provides synchronization of the data between the different users' applications using Web services. Data synchronization is done only when it is necessary and the connection to the server is available.

Only the parent and the teacher can set up synchronization, while the pupil (child) can only use it. The renovation of the educational content can be done both by the teacher and the parent. In addition, they can upload media objects, define daily or monthly tasks, as well as the expected objectives. If the teacher does not agree with the content uploaded by the parent, he/she can remove it.

Only the teacher can assess the results achieved by the children, while parents can only see the assessment, but cannot change it.

The children can access and utilize various resources: textual and multimedia information, tasks, and view their grades.

The parent may monitor the complete diary of the child who uses the application; the duration of working with it; if a new version of the application is available, the parent can install it, being authorized by their account in the application of the child. In this way the child's access the Internet is limited.

Also, it should be pointed out that in most cases the assessment is stressful for preschool children. That is why in this application the term for score is 'award' or other stimulus that provokes the child to strive for higher achievements.

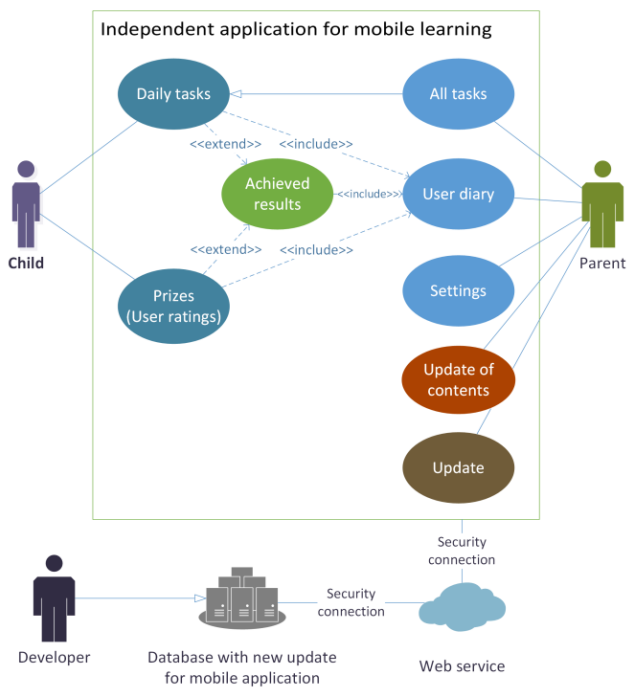

Figure 1. A conceptual model of a mobile application for individual learning by young children.

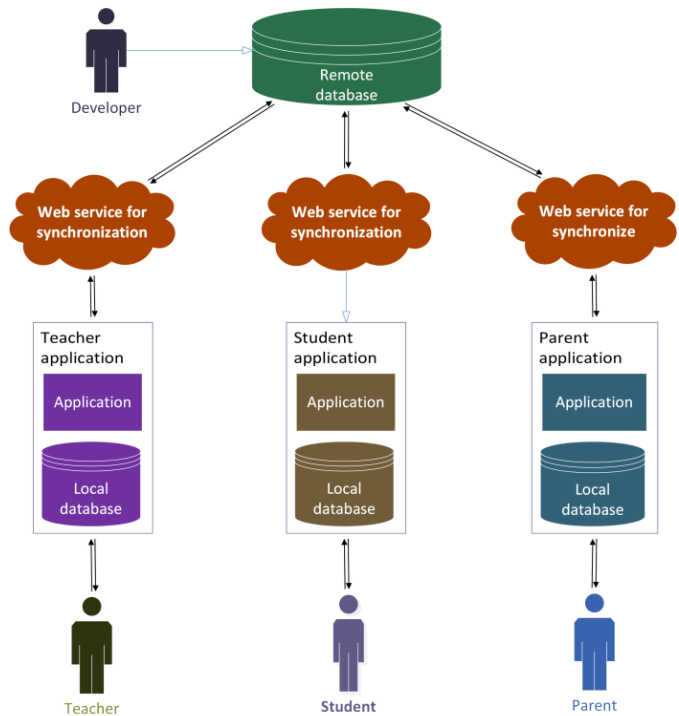

Figure 2. A conceptual model of a mobile application for collective learning.

Some of the advantages of using the proposed conceptual model for collective mobile learning are given below. The major one is that the parents can actively participate in the education of their children. Another advantage is the provided safe environment to work when using a mobile device in terms of limiting access to the Internet.

The disadvantage of this model is the complexity of software development.

## VII. CONCLUSION

Undoubtedly the design and quality application development for mobile learning is an important part of the software technology future.

The educational content requirements stated in SER for preschoolers in Bulgaria have been presented in this article. The issue of safety of children using mobile devices relating to the protection of their personal data on the Internet has been discussed. The criteria for assessing applications for mobile learning of young children have been identified.

As a result of this study two conceptual models for mobile learning have been proposed.

The frameworks for the mobile learning presented here are based on continuous study of best practices in this area. The proposed conceptual models have been illustrated by diagrams. Their positive and negative characteristics have been pointed out.

The practical developments of the mobile learning frameworks are to be made. They are going to be the subject of further research in view of their practicality and applicability.

REFERENCES

[1] M. Sarwar, T. R. Soomro, "Impact of smartphone's on society", European journal of scientific research, vol. 98, no. 2, pp. 216-226, 2013

[2] T. M. Cumming, "Sustaining mobile learning: Theory, research, and practice", edition: 1st, publisher: Routledge, editor: Ng, W. & Cumming, T., ISBN: 978-1-138-78738-4, 2015

[3] D. Donchev, I. Hristov, "Mobile learning applications ubiquitous characteristics and technological solutions", Journal of cybernetics and information technologies, vol. 6, no 3, Sofia, 2006

[4] G. Botzer, M. Yerushalmy, "Mobile application for mobile learning", IADIS International Conference on Cognition and Exploratory Learning in Digital Age (CELDA 2007), Portugal, 2007

[5] D. Parsons, H. Ryu, "A framework for assessing the quality of mobile learning", in proceedings of the 11th International Conference for Process Improvement, Research and Education (INSPIRE), Southampton Solent University, UK, 13 April 2006

[6] J. S. Eccles, P. E. Davis-Kean, "Influences of parents' education on their children's educational attainments: the role of parent and child perceptions", London review of education, vol. 3 (3), pp. 191-204, 2005

[7] R. Kraleva, A. Stoimenovski, D. Kostadinova, V. Kralev, "In-vestigating the opportunities of using mobile learning by young children in Bulgaria", International journal of computer science and information security (IJCSIS), vol. 14, no 4, April 2016 (in appear)

[8] Z. Kalinić, S. Arsovski, "Mobile learning – quality standards, requirements and constraints", International journal for quality research, vol. 3, no. 1, pp. 7-17, 2009

[9] Ministry of Education, "Decree № 4 of 18.09.2000 on pre-school education" (in Bulgarian), http://www.mon.bg/?h=downloadFile&fileId=145, [accessed January 2016]

[10] S. Livingstone, B. O'Neill, "Children's rights online: challenges, dilemmas and emerging directions", In Van Der Hof, S., Van Den Berg, B., and Schermer, B. (eds), Minding minors wandering the web: Regulating online child safety, pp.19-38, Berlin: Springer, 2014

[11] S. Livingstone, J. Carr, J. Byrne, "One in three: Internet governance and children's rights", Journal global commission on internet governance, paper series no. 22, Nov. 2015

[12] M. Sarwar, T. R. Soomro, "Impact of Smartphone's on Society", European journal of scientific research, vol. 98 (2), p. 216-226, 2013

[13] J. S. Radesky, J. Schumacher, B. Zuckerman, "Mobile and interactive media use by young children: the good, the bad, and the unknown", PEDIATRICS, vol. 135 (1), DOI: 10.1542/peds.2014-2251, 2014

[14] B. Scifo, "Investigating the domestication of convergent mobile media and mobile internet by children and teens: preliminary issues and empirical findings on opportunities and risks", Open journal obra digital, vol. 4, 2014

[15] A. Ali, A. Ouda, L. F. Capretz, "A conceptual framework for measuring the quality aspects of mobile learning", Bulletin of the IEEE Technical committee on learning technology, vol. 14, no 4, Oct. 2012

[16] T. Elias, "71. Universal instructional design principles for mobile learning", International review of research in open and distance learning, vol. 12 (2), Febr. 2011

[17] A. Stoimenovski, R. Kraleva, V. Kralev, "Analysis of mobile applications suitable for mobile learning of preschool children", First student and PhD scientific session (SDSS-2016), South-West University "Neofit Rilski", Blagoevgrad, 19-20 May 2016 (in appear)

[18] R. Kraleva, "Acoustic-phonetic modelling for children's speech recognition in Bulgaria", PhD thesis, South-West University, Blagoevgrad, Bulgaria, 2014

[19] Z. L. Berge, L. Y. Muilenburg, "Handbook of mobile learning", Taylor & Francis Group, 2013

[20] W. Ng, T. M. Cumming, "Sustaining mobile learning: Theory, research and practice", Taylor & Francis Group, 2016

[21] A. Druin, "Mobile technology for children: Designing for interaction and learning", Elsevier Inc., 2009

[22] G. Vavoula, C. Karagiannidis, "Designing Mobile Learning Experiences", Advances in informatics 2005, ISBN: 978-3-540-29673-7 (Print), 2005

[23] Y. C. Hsu, Y. H. Ching, "A review of models and frameworks for designing mobile learning experiences and environments", Canadian journal of learning & technology, vol. 41 (3), 2015

[24] Learning and skills development agency, "Learning with mobile devices research and development", Book of papers edited by Jill Attewell and Carol Savill-Smith, Learning and Skills Development Agency, 2004